\documentclass[fleqn,usenatbib]{mnras}

\usepackage{newtxtext,newtxmath}

\usepackage[T1]{fontenc}

\DeclareRobustCommand{\VAN}[3]{#2}
\let\VANthebibliography\thebibliography
\def\thebibliography{\DeclareRobustCommand{\VAN}[3]{##3}\VANthebibliography}

\usepackage{graphicx}	
\usepackage{amsmath}	
\usepackage{xcolor,xspace}
\usepackage{orcidlink}
\usepackage{threeparttable}

\newcommand{\nustar}{{\it NuSTAR}\xspace}
\newcommand{\swift}{{\it Swift}\xspace}
\newcommand{\xmm}{{\it XMM-Newton}\xspace}
\newcommand{\SRG}{{\it SRG}\xspace}
\newcommand{\ero}{\mbox{eROSITA}\xspace}

\newcommand{\xspec}{\texttt{XSPEC}\xspace}
\newcommand{\grppha}{\texttt{grppha}\xspace}

\newcommand{\nupip}{\texttt{nupipeline}\xspace}
\newcommand{\nupro}{\texttt{nuproducts}\xspace}
\newcommand{\barycorr}{\texttt{barycorr}\xspace}
\newcommand{\efsearch}{\texttt{efsearch}\xspace}

\newcommand{\xrtpip}{\texttt{xrtpipeline}\xspace}
\newcommand{\bxa}{\texttt{BXA}\xspace}

\newcommand{\ergcm}[1]{$\times 10^{#1}$ erg cm$^{-2}$ s$^{-1}$}
\newcommand{\ergs}[1]{$\times 10^{#1}$ erg s$^{-1}$}
\newcommand{\uergcms}{erg cm$^{-2}$ s$^{-1}$}
\newcommand{\uergs}{erg s$^{-1}$}

\newcommand{\src}{eRASSU\,J012422.9$-$724248\xspace}
\newcommand{\osrc}{SMC732.10.7\xspace}
\newcommand{\angstrom}{\mbox{\normalfont\AA}\xspace}

\usepackage{ulem} 

\title[Be/X-ray binary pulsar eRASSU\,J012422.9$-$724248 in the Magellanic Bridge]{Broadband study of the Be/X-ray binary pulsar eRASSU\,J012422.9$-$724248 in the Magellanic Bridge, near the Eastern Wing of the Small Magellanic Cloud}

\author[Haonan Yang et al.]{
Haonan Yang\orcidlink{0000-0002-7680-2056},$^{1,2}$\thanks{Corresponding: hnyang@nao.cas.cn}
Chandreyee Maitra\orcidlink{0000-0002-0766-7313},$^{3,2}$\thanks{Corresponding: cmaitra@mpe.mpg.de}
Frank Haberl\orcidlink{0000-0002-0107-5237},$^{2}$
David Kaltenbrunner\orcidlink{0000-0003-3645-2853},$^{2}$
Lorenzo Ducci\orcidlink{0000-0002-9989-538X},$^{4}$ \newauthor
Andrzej Udalski\orcidlink{0000-0001-5207-5619},$^{5}$
and Georgios Vasilopoulos\orcidlink{0000-0003-3902-3915}$^{6,7}$
\\
$^{1}$National Astronomical Observatories, Chinese Academy of Sciences, 20A Datun Road, Beijing 100101, China\\
$^{2}$Max-Planck-Institut f\"{u}r extraterrestrische Physik, Gie\ss{}enbachstra\ss{}e 1, D-85748 Garching bei M\"{u}nchen, Germany\\
$^{3}$Inter-University Centre for Astronomy and Astrophysics (IUCAA), Ganeshkhind, Pune 411007, India\\
$^{4}$Institut f\"{u}r Astronomie und Astrophysik, Universit\"{a}t T\"{u}bingen, Sand 1, 72076 T\"{u}bingen, Germany\\
$^{5}$Astronomical Observatory, University of Warsaw, Al. Ujazdowskie 4, 00-478 Warszawa, Poland\\
$^{6}$Department of Physics, National and Kapodistrian University of Athens, University Campus Zografos, GR 15784, Athens, Greece\\
$^{7}$Institute of Accelerating Systems \& Applications, University Campus Zografos, GR 15784, Athens, Greece
}

\date{Accepted XXX. Received YYY; in original form ZZZ}

\pubyear{2025}

\begin{document}
\label{firstpage}
\pagerange{\pageref{firstpage}--\pageref{lastpage}}
\maketitle

\begin{abstract}
The first four all-sky surveys with \ero the soft X-ray instrument on board the Spektrum-Roentgen-Gamma (SRG) satellite revealed a new X-ray source, \src, in the Magellanic Bridge, near the Eastern Wing of the Small Magellanic Cloud (SMC). We performed a broadband timing and spectral analysis using the optical and X-ray data of \src. Using the X-ray observations with \ero, \swift, \nustar and optical data from the optical Gravitational Lensing Experiment (OGLE) and the Las Cumbres Observatory (LCO), we confirm the nature of \src as a Be/X-ray binary (BeXRB) pulsar in the Magellanic bridge.
The position is coincident with that of an early-type star (OGLE ID \osrc). We detect the spin period at 341.71 s in \nustar data and infer a period of 63.65 days from the 15 year monitoring with OGLE, that we interpret as the orbital period of the system. A tentative CRSF at $\sim$12.3 keV is identified in \nustar spectra with $\sim 1.8\sigma$. The source appears to show a persistent X-ray luminosity and an optical magnitude transition on the long timescale. We propose \src is a new member of the class of persistent BeXRBs.
\end{abstract}




\section{Introduction}

The nearest star-forming galaxies, the Magellanic Clouds (MCs), are known to host a large population of high-mass X-ray binaries \citep[HMXBs;][]{2016MNRAS.459..528A, 2016A&A...586A..81H}. The accurately known distances \citep[][]{2014ApJ...780...59G, 2019Natur.567..200P} and relatively low foreground absorption toward the MCs \citep[][]{2024MNRAS.534.3478N} make them ideal targets to study the properties of HMXBs in detail. Notably, the majority of the HMXBs in the MCs are revealed to be Be/X-ray binaries (BeXRBs), a subclass in which typically a neutron star accretes matter from a Be-type star (see \citealt{2011Ap&SS.332....1R} for a review).
These systems typically show high variability and are often detected during outburst phases. This includes Type I outbursts, occurring at the periastron passage of the neutron star, and Type II outbursts, which are more intense and often associated with significant changes in the circumstellar disk of the Be star \citep[e.g.][]{2020MNRAS.494.5350V, 2025MNRAS.536.1357Y}.  Cyclotron resonance scattering features (CRSFs) were detected in several BeXRBs, which  indicate a strong magnetic field \citep{2017JApA...38...50M,2019A&A...622A..61S}.

Since its launch in 2019, the \ero instrument onboard the {\it Spectrum-Roentgen-Gamma} (\SRG) observatory \citep{2021A&A...647A...1P, 2021A&A...656A.132S} has provided a notable enhancement in the knowledge of X-ray sources. 
During the four all-sky surveys conducted between 2020-2022, \ero monitored the X-ray population of the entire Magellanic system which includes the Large and Small Magellanic Clouds (LMC and SMC), and the Magellanic Bridge connecting them. 
The Magellanic bridge extends across 200 degrees on the sky, and consists of neutral gas and a stellar population with indications of a younger population \citep{1990AJ.....99..191I, 2002ApJ...578..126L, 2010ApJ...723.1618N}. Tidal interactions between the LMC and SMC 200 Myr ago are suggested to be the primary formation mechanism, pulling material from the SMC Wing during a close encounter \citep{2004ApJ...616..845M, 2015ApJ...813..110H}.
Several transient HMXBs were identified in the Magellanic Bridge with a denser spatial distribution toward the SMC. They correlate with younger stellar populations, which suggests an origin in a tidally induced star-forming episode \citep{2005A&A...435....9K, 2010MNRAS.403..709M, 2014MNRAS.444.3571S, 2020MNRAS.495.2664C, 2020MNRAS.494.1424C}.

During the first four all-sky surveys with \ero (eRASS1$-$4), \src was discovered in the Magellanic Bridge and identified as a new BeXRB \citep{2023ATel15886....1M}. Follow-up Directors Discretionary Time (DDT) observations with the Nuclear Spectroscopic Telescope ARray \citep[\nustar;][]{2013ApJ...770..103H} revealed coherent X-ray pulsations at $\sim$341.7~s, which further confirm it as a neutron star BeXRB. 

This paper is organized as follows. We describe the X-ray and optical observations of \src in Section~\ref{sec:obs}. In Section~\ref{sec:analysis} we detail the identification of the optical counterpart and X-ray properties derived from various data sets, including spectral and timing analysis. We discuss the results in Section~\ref{sec:discuss}.

\section{Observations and data reduction}\label{sec:obs}

\subsection{\ero}

The eRASS surveys scanned the entire sky through a series of great circles intersecting at the ecliptic poles between December 2019 to February 2022. Each \ero survey (eRASSn) recurs at six-month intervals, enabling multi-year monitoring of the X-ray source populations in the Magellanic system.

\src was discovered after it was scanned for a total exposure time of 1.3 ks (after vignetting corrections) during eRASS1$-$4.
We extracted \ero\ data products using the \ero\ Standard Analysis Software System \citep[\texttt{eSASS} version \texttt{eSASSusers\_211214};][]{2022A&A...661A...1B, 2024yCat..36820034M}.
{We extracted} the source and background events using the \texttt{eSASS} task \texttt{srctool}.
For the light curve, we combined the data from all telescope modules (TM 1--7) and applied a cut on fractional exposure of 0.15. For the spectral analysis, we combined the data from TMs with an on-chip optical block filter (TM 1--4 and 6). TM5 and TM7 were not used because no reliable energy calibration is available so far due to the optical light leak \citep{2021A&A...647A...1P}.

\subsection{\swift}
Following the discovery of \src with \ero, we performed a follow-up observation on 2023 January 15 between 18:12 and 23:08 UTC using the X-ray telescope \citep[XRT;][]{2004ApJ...611.1005G, 2005SSRv..120..165B} onboard the {\it Neil Gehrels Swift observatory} \citep[\swift;][]{2004ApJ...611.1005G} (MJD 59959.76 - 59959.96, ObsID 00015841001). The 2.2 ks exposure was taken in Photon Counting (PC) mode given the relatively low X-ray flux according to the \ero monitoring. We used the standard \xrtpip version 0.13.7 and CALDB version 20240506 for the \swift/XRT data reduction. The source was detected with an XRT count rate of 2.01$\pm$0.33 $\times$ 10$^{-2}$ cts s$^{-1}$ in the 0.3$-$10.0 keV band. Based on the detected position, we used a circle with a radius of 20\arcsec\ as the source extraction region and defined the background region as a nearby circle with a radius of 50\arcsec.

\subsection{\nustar}
After the discovery, \nustar observed \src on 2023 January 22 (MJD 59966.44 - 59967.02, ObsID 90901301002) and January 23 (MJD 59967.58 - 59968.43, ObsID 90901301004) for a total effective exposure of 75.3 ks on FPMA and 75.8 ks on FPMB, respectively (see Table \ref{tab:log} for a summary).
We reduced the \nustar data with \nupip version 0.4.9 and CALDB version 20200425. 
For both source and background regions, we used circular regions with radii of 31.9\arcsec.
Then we extracted spectra and light curves of FPMA and FPMB instruments with \nupro version 0.3.1. 
Photon arrival times were corrected to the equivalent time at the solar system barycenter using the \barycorr tool.

\begin{table}
\caption{
\label{tab:log}
Journal of the 2023 X-ray observations of \src\ presented in this work.}
\centering
\begin{tabular}{lcccc}
\hline\hline
Instrument	& Obs ID 		&Start time (UTC)	   & Exposure  \\
	        &			    & Mmm DD hh:mm:ss	   & (ks)    \\
\hline
\swift/XRT                  & 00015841001	  & Jan 15 18:06:36   & 2.2 \\
\nustar                     & 90901301002     & Jan 22 10:16:11   & 32.6 \\
\nustar                     & 90901301004     & Jan 23 13:36:12   & 44.4 \\

\hline
\hline
\end{tabular}
\newline
{\bf Notes.} The \swift was operated in Photon Counting (PC) mode.
\end{table}

\subsection{OGLE}
\label{sec:ogle}

The field around \src was observed by the Optical Gravitational Lensing Experiment \citep[OGLE;][]{1992AcA....42..253U}. 
The optical counterpart, a blue star listed with V = 14.8\,mag in the UCAC4 catalogue \citep{2013AJ....145...44Z} was monitored in the I band during the phases III \citep[OGLE-III ID: SMC120.3.3254,][]{2014A&A...562A.125K} and IV \citep[OGLE-IV ID: \osrc,][]{2015AcA....65....1U} of the OGLE project.
Images were taken at the Las Campanas Observatory in the $I$-band with the 1.3\,m Warsaw telescope and the magnitudes were calibrated to the standard $I$-band system as described in \citet{2015AcA....65....1U}.
The OGLE-III light curve was published by \citet{2014A&A...562A.125K} and is available from CDS\footnote{\url{https://vizier.cds.unistra.fr/viz-bin/VizieR}}.

\subsection{LCO}
\label{sec:lco}

We used the optical spectroscopic data from the Folded Low Order whYte-pupil Double-dispersed Spectrograph (FLOYDS) on the Las Cumbres Observatory (LCO) 2-m telescope at Siding Spring Observatory \citep{2013PASP..125.1031B} to characterize the optical counterpart of \src.
The counterpart was identified with the Gaia DR3 source Gaia DR3 4687297547582926464, which was used for telescope pointing. The observation started on 2022 November 18 (MJD 59901.56) with an exposure time of 1.3 ks. The spectrum has a wavelength range of 3200--10000 \AA\ with a spectral resolution of R$\sim$400--700. We used the red/blue grism and set the slit width to 2 arcsec.
As part of the FLOYDS pipeline\footnote{\url{https://lco.global/documentation/data/floyds-pipeline/}}, \texttt{PyRAF} tasks were used to reduce the spectrum. We paid attention on the H$\alpha$ and H$\beta$ lines, which typically appear in the spectra of Be stars \citep{1988PASP..100..770S, 2000ASPC..214....1B, 2003PASP..115.1153P}.

\section{Analysis}\label{sec:analysis}

\subsection{X-ray position and optical counterpart}

\src was detected by \ero at the position (after astrometrical corrections) of $\alpha_\mathrm{J2000.0}=01^\mathrm{h}24^\mathrm{m}22^\mathrm{s}.9$ and $\delta_\mathrm{J2000.0}=-72^\circ 42'48''.7$ with a statistical uncertainty of $1.4''$.
Within the X-ray position error circle, we identified an early-type star, suggesting the BeXRB nature of \src.

\begin{figure}
\centering
\includegraphics[width=1.0\columnwidth]{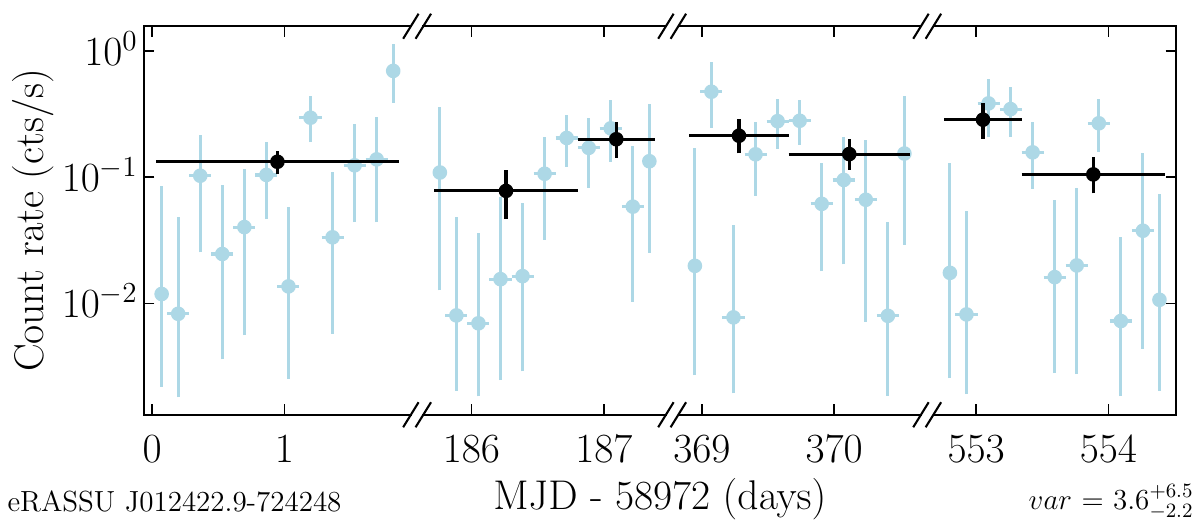}
\caption{\ero light curve of \src, where the blue points represent time binning per \ero scan and the black points show results in which each bin contains a minimum of ten counts to improve statistics. Error bars in the $x$-direction give the length of time bins. For the bins with fewer than ten counts, the 1-$\sigma$ upper limits are plotted.}\label{fig:erosita_lc}%
\end{figure}

\subsection{X-ray timing analysis}

\subsubsection{The \ero light curve}

Figure~\ref{fig:erosita_lc} shows the \ero X-ray light curve of \src in the energy range of 0.2--5 keV from all four \ero all-sky surveys (eRASS1--4). We initially extracted the light curve with \texttt{srctool} in 1~s bins, then the light curve shown in Figure~\ref{fig:erosita_lc} was created by fitting the observed counts in the source and background regions using Bayesian inference. We assumed the counts to follow a Poisson distribution and that within each time bin, the count rate was constant. These assumptions were necessary due to the changing fractional exposure caused by \ero's scanning scheme. We created one light curve binned per \ero scan ($\sim$40~s) and one to contain a minimum of 10 net source counts per time bin for better statistics.
Similar to \citet{2008A&A...480..599S, 2013A&A...558A...3S}, the variability $var$ was defined as $var=src_\mathrm{max}/src_\mathrm{min}$, where $src$ and $\sigma$ are the source count rates and corresponding uncertainties at the bins where $src-\sigma$ has its maximum and where $src+\sigma$ has its minimum (subscripts 'max' and 'min', respectively).
The count rate of \src stays at a similar level during the four survey intervals, with the variability of $var=3.6_{-2.2}^{+6.5}$.

\subsubsection{Searching for variability and pulsations}
\begin{figure}
\centering
\includegraphics[width=1.0\columnwidth]{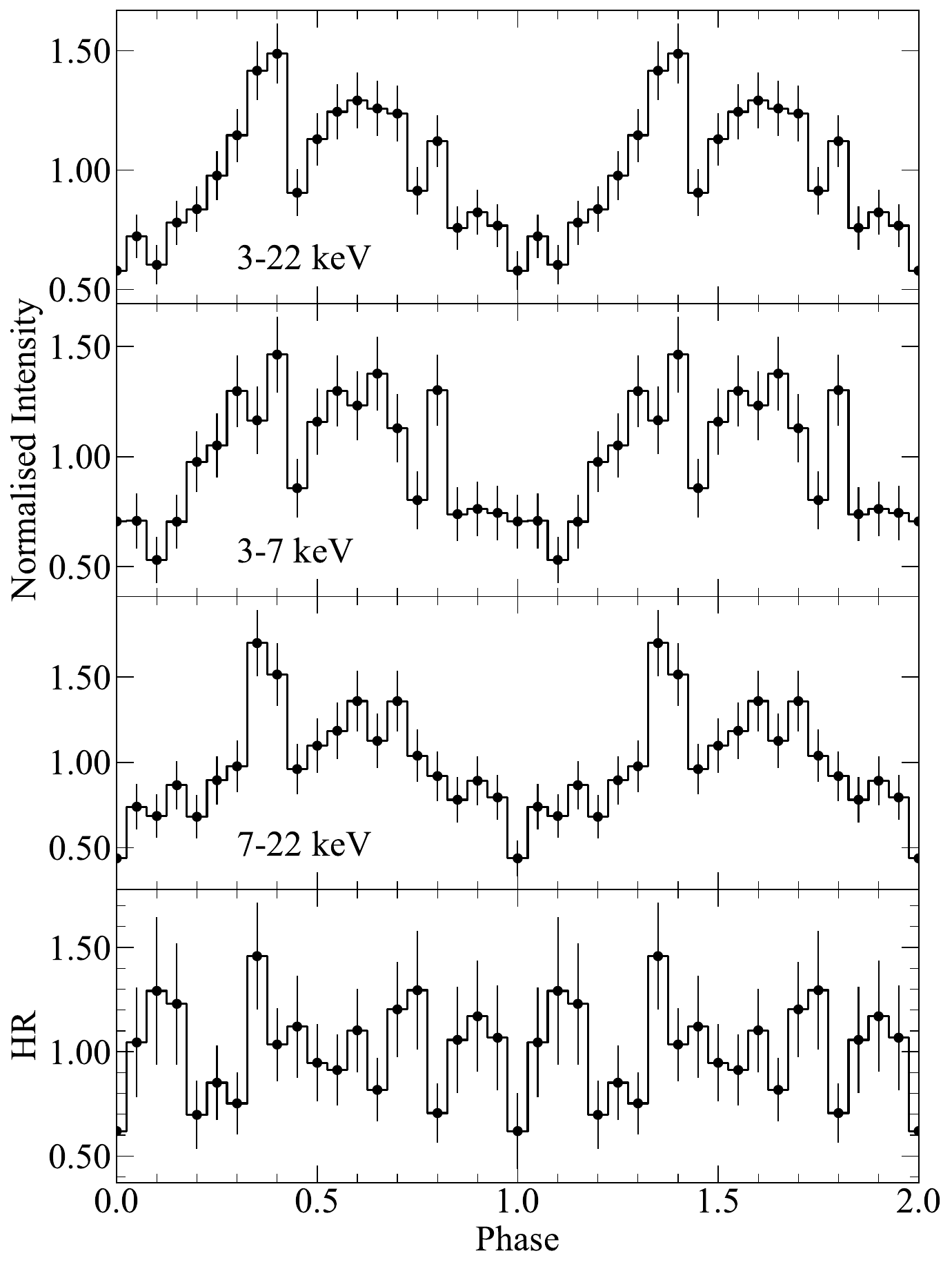}
\caption{\nustar pulse profiles obtained from folding the combined light curves from two observations in different energy bands and the hardness ratio. There is no statistically significant difference between the individual pulse profiles of the two observations.
}
\label{fig:pulse_profile_nustar}
\end{figure}

We used the barycenter-corrected event files of the two \nustar observations to generate light curves with a time bin of 0.01s for FPMA and FPMB. Given that the energy bands above $\sim$22.0 keV are dominated by background photons, we extracted light curves within 3--22 keV. 
The average background-subtracted source count rates in this band were 0.04 ct/s for both FPMA and FPMB of the two observations.
For each observation, we combined background-subtracted light curves from both FPMA and FPMB modules. Then we concatenated the light curves from the two observations and searched for periodic signals using a Lomb-Scargle periodogram analysis \citep{1976Ap&SS..39..447L, 1982ApJ...263..835S}. We found a significant signal at $\sim$341 s, which is also detected using the \efsearch tool. To estimate the uncertainty of the period, we generated a series of 1000 light curves following the method of \citet{1999ApJ...522L..49G} for both \nustar observations and combined them in the same way as we do for the observation data. The standard deviation of the distribution of periods measured from the simulated light curves was used as the 1-$\sigma$ uncertainty. This results in a period of 341.71$\pm$0.04\,s. We folded the light curve with this period with phase 0 defined at MJD 59966.6. 
Based on the count distribution, we split the 3--22 keV band at 7 keV into two. This provides a similar number of counts below and above 7 keV (56\% and 44\%). The pulse profiles in the different energy bands are shown in Figure~\ref{fig:pulse_profile_nustar}, which are characterized by a multi-peak structure. 
The primary and secondary peaks at phases $\sim$0.4 and $\sim$0.6 are visible in both soft and hard bands, while a third weaker and narrower peak at phase $\sim$0.8 is only significant in the 3--7 keV profile.

\begin{figure}
\centering
\includegraphics[width=1.0\columnwidth]{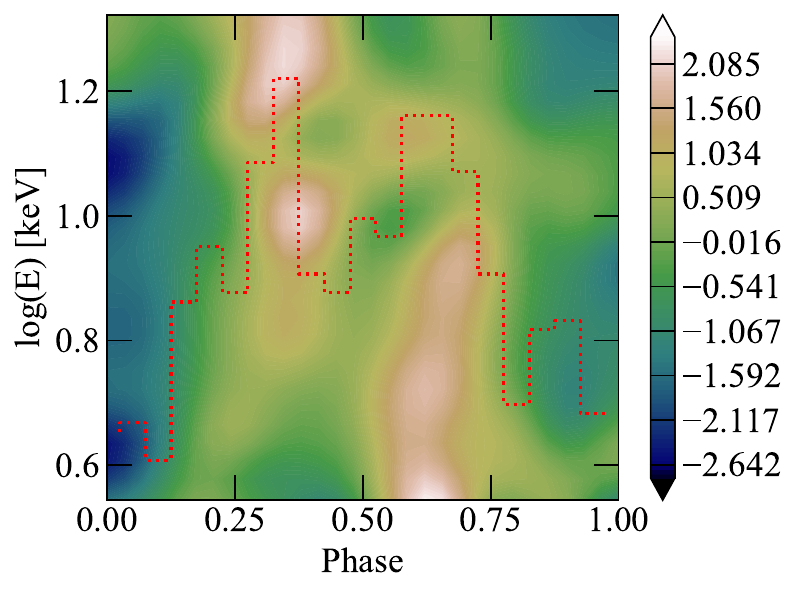}\\
\includegraphics[width=1.0\columnwidth]{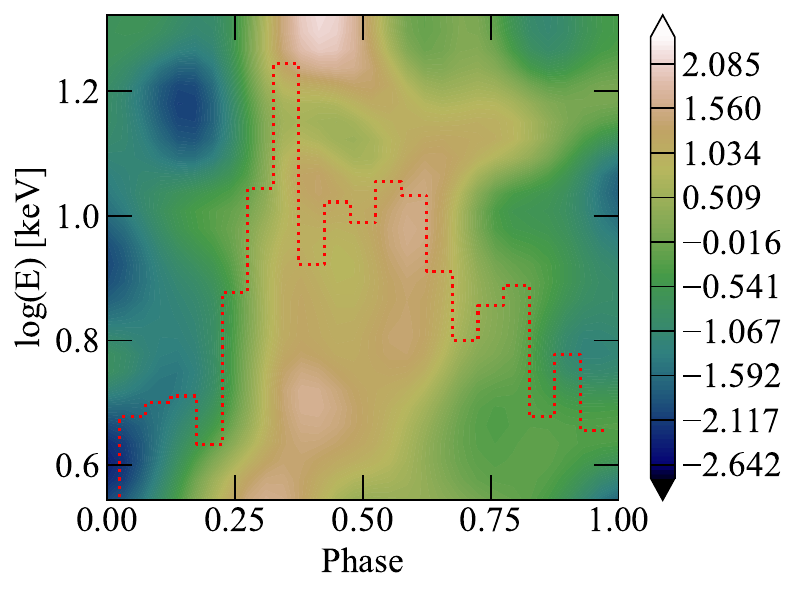}
\caption{Upper panel: \nustar phase energy heat-map obtained from the combined light curve from observation 90901301002. Lower panel: The heat-map obtained in the same way from observation 90901301004.
Each energy bin is normalised by subtracting the average pulse intensity and subtracted by the standard deviation of the energy bin. The 3-22 keV pulsed profile is plotted as red dashed histogram.}\label{fig:heat_both}
\end{figure}

\begin{figure}
\centering
\includegraphics[width=1.0\columnwidth]{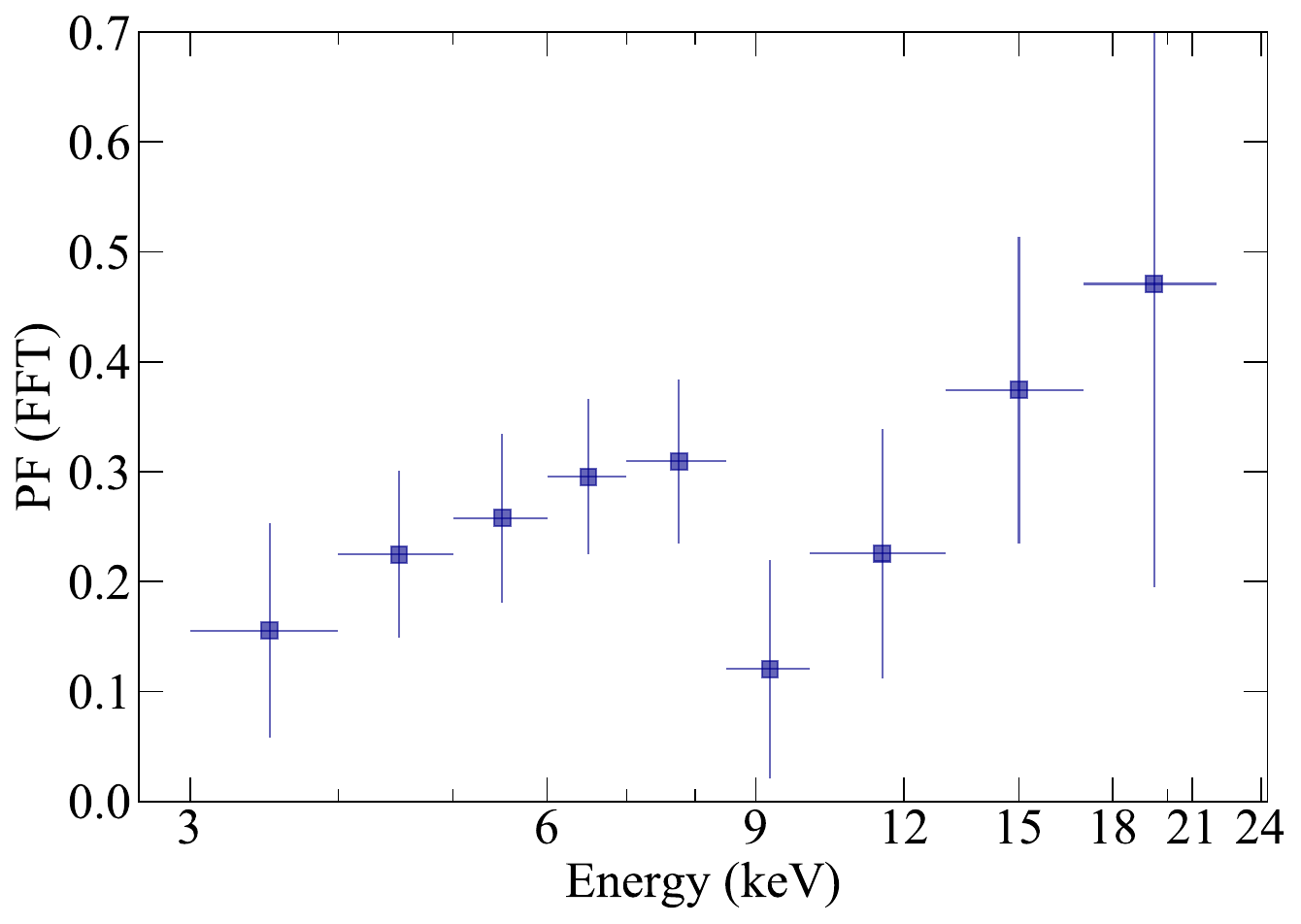}
\caption{\nustar pulsed fraction as a function of energy. The combined data of two observations are used, which reveals a drop at around 10 keV. Such a feature could not be significantly constrained using data from either observation alone due to low statistics.
}\label{fig:pulsed_fraction}
\end{figure}

We also examined the pulse profiles from the two observations individually. The differences between the pulse profiles from the two observations were not statistically significant. However, when we plotted the normalised pulse profiles as a function of energy in the form of heat-maps, as shown in Figure~\ref{fig:heat_both}, 
the two observations seem to indicate slightly different energy evolutions especially with respect to the 10\,keV feature. This cannot be ascertained in confidence due to the statistical limitations of the data.
We estimated the pulsed fraction, which refers to the magnitude of the pulsed component relative to the total emission, over the 3--22 keV band using the combined light curve of two observations following the fast Fourier transform method of \citet{2023A&A...677A.103F} (see Figure~\ref{fig:pulsed_fraction}). 
We notice a drop at around 10 keV, which could be related to the CRSF feature in the spectrum as described in Section~\ref{sec:spec_nustar}. The energy-dependent trend was consistently recovered with the independent root mean square (rms) method.
Such a change in the morphology of the pulsed fraction, however, could not be significantly constrained using either observation alone.

\begin{figure}
\centering
\includegraphics[width=1.0\columnwidth]{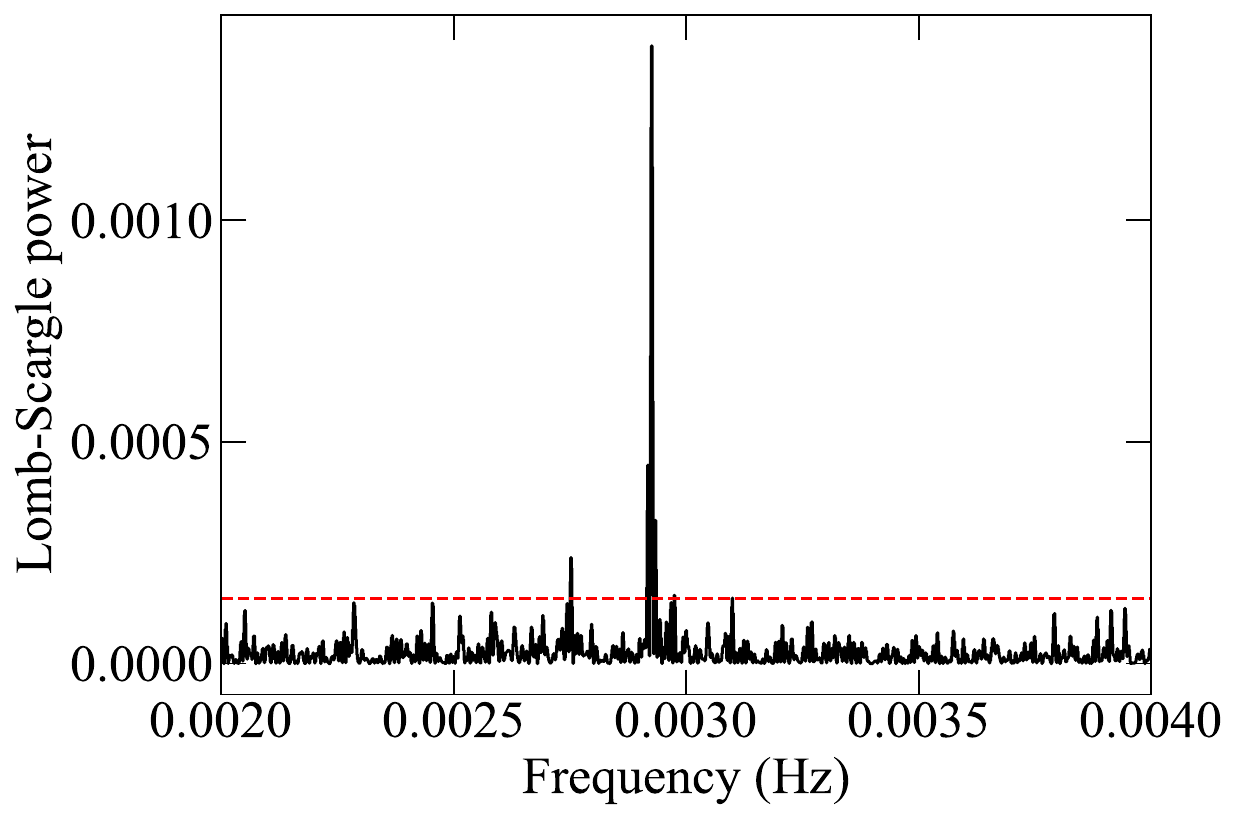}
\caption{The Lomb-Scargle periodogram of \src obtained from the combined \nustar data (3.0-22.0 keV). The dashed red line marks the 99.73\% (3-$\sigma$) confidence level obtained from the simulation.}\label{fig:psd_nustar}
\end{figure}

\subsection{X-ray spectral analysis}

For the analysis of the X-ray spectra we used \xspec version 12.14.1 \citep{1996ASPC..101...17A}. For the Galactic absorption $N_\mathrm{H}$ we used \texttt{tbabs} and fixed the value at $5.4\times 10^{20}$ cm$^{-2}$ from the NH calculator using \citet{1990ARA&A..28..215D} with a correction factor of 1.25 to account for molecular Hydrogen (following the minimum expected amount according to \citealt{2013MNRAS.431..394W}). We adopted abundances from \citet{2000ApJ...542..914W} and cross-sections from \citet{1996ApJ...465..487V}. For the distance we assumed 60 kpc.

\subsubsection{\ero}
\label{sec:ero_spec}

We used the Bayesian X-ray analysis (\bxa) software \citep{2014A&A...564A.125B} to perform the spectral analysis. \bxa connects the X-ray spectral analysis environment \xspec to the nested sampling algorithm UltraNest \citep{2021JOSS....6.3001B}, allowing us to explore the entire model parameter space with Bayesian parameter estimation.

We modelled the spectrum of \src with an absorbed power law. 
We fitted the source and background spectra simultaneously with the principal component analysis (PCA) background model provided by \bxa. The shape of the PCA model came from the spectrum of the off-region. We then fitted the normalisation of the shape together with the source model while tying the normalisations of the background components from the two regions to one another with the \texttt{BACKSCAL} values (calculated by \texttt{srctool}) used as scaling factor.
An additional \texttt{tbvarabs} component with metallicity of 0.2 \citep{1998AJ....115..605L} did not improve the fit significantly.

The spectrum is best fit with a power-law index of $0.58^{+0.29}_{-0.30}$ and a flux of 3.9$^{+1.0}_{-0.9}$ \ergcm{-13} (0.2--5\,keV), which corresponds to an absorption-corrected luminosity of 1.6$^{+0.4}_{-0.3}$ \ergs{35}.

\begin{figure}
\centering
\includegraphics[width=1.0\columnwidth]{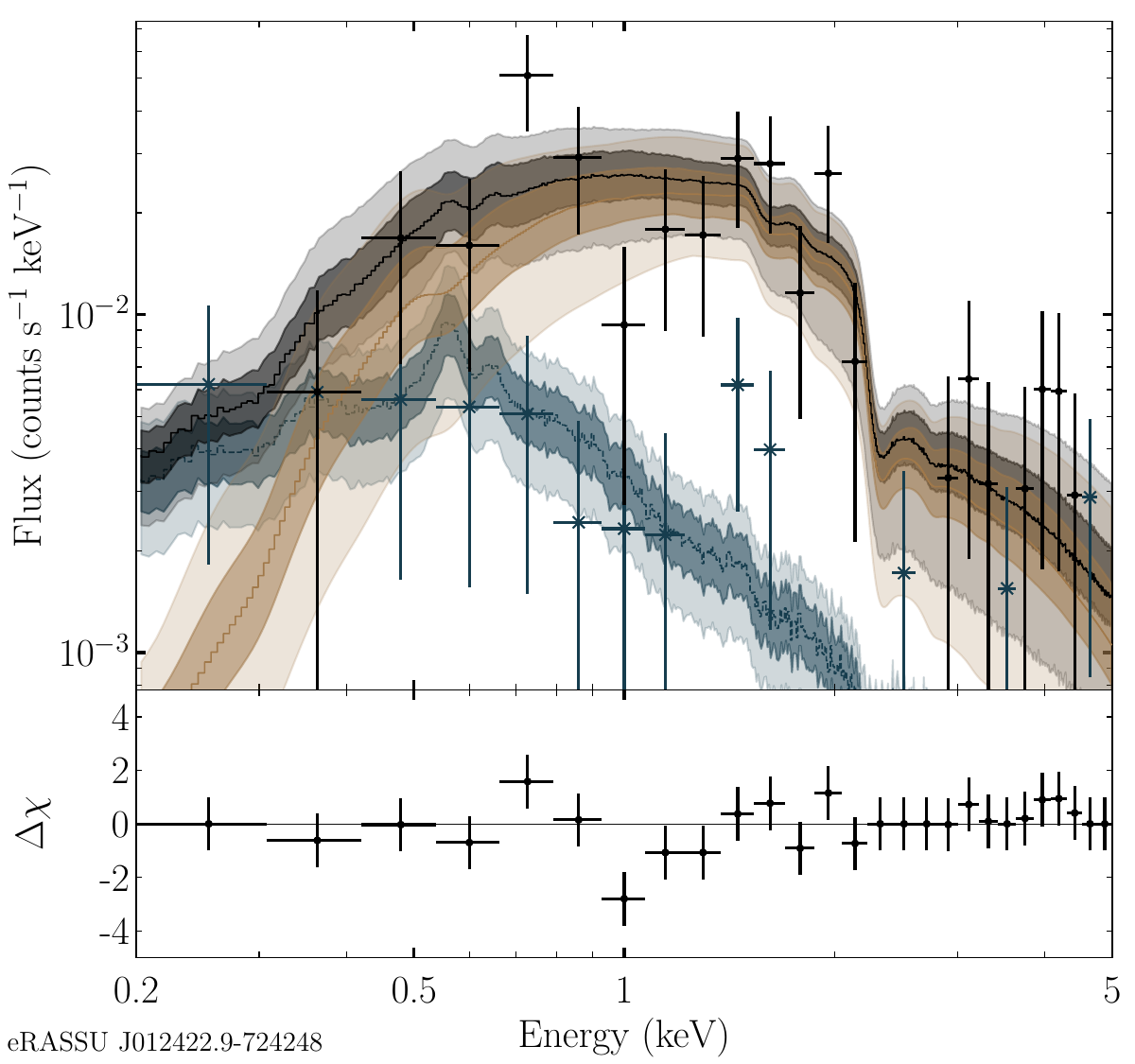}
\caption{Upper panel: The \ero spectra of \src with best-fit model. Data from the source and background are represented by points and crosses, respectively. The source and background models are shown as solid dark yellow lines and dashed dark blue lines, while the total model is shown as a solid black line. Lower panel: The corresponding residuals for the total model.}\label{fig:spec_ero}
\end{figure}

\subsubsection{\swift}
We binned the \swift/XRT spectra to achieve a minimum of one count per bin using \grppha, which is suitable for the C-statistic.
The spectrum could be described with an absorbed power-law model. For the Galactic absorption $N_\mathrm{H}$ we used the \texttt{tbabs} model and fixed it at the same value as the \ero data. The measured photon index was $\Gamma=0.2\pm0.5$ and the observed absorption-corrected 0.3--10 keV flux was $2.6_{-1.0}^{+1.5}\times 10^{-12}$ \uergcms. Errors were estimated at 90\% confidence intervals. The corresponding X-ray luminosity is $1.1\times 10^{36}$ \uergs.

\subsubsection{\nustar}\label{sec:spec_nustar}

We rebinned and grouped the data of two observations using \grppha to ensure at least one count per bin, allowing us to use C-statistic to evaluate the fit. 
We examined the relative intensities of the source and background spectra in \xspec, and removed the background-dominated bins of the data, including the bands above 22.0 keV.

We jointly fitted the FPMA and FPMB spectra from the two \nustar observations in the 3--22 keV energy band using an absorbed cutoff power-law model (\texttt{cutoffpl}) incorporating a cross-normalisation \texttt{constant} component. We fixed The cross-normalisation constant at unity for FPMA in the first observation (ObsID 90901301002), while allowing that for FPMB in the first observation and both modules in the second observation (ObsID 90901301004) to vary freely. This basic model provided a statistically acceptable fit to the observed spectra, with cstat = 711.60 for 939 degrees of freedom. 
A further test with several alternative continuum models including a power law with a high-energy cutoff (\texttt{powerlaw} * \texttt{highecut}) and a thermally Comptonized continuum (\texttt{nthcomp}) did not yield a statistically better fit. Furthermore, the \texttt{compTT} + \texttt{compTT} model which has been used in some low-luminosity BeXRB systems \citep[e.g.][]{2012A&A...540L...1D, 2019MNRAS.483L.144T} did not improve the fit given the limited counts and the exclusion of the high-energy band above 22 keV. A single \texttt{comptt} component also failed to provide a better description of the data.

\begin{figure}
\centering
\includegraphics[width=1.0\columnwidth]{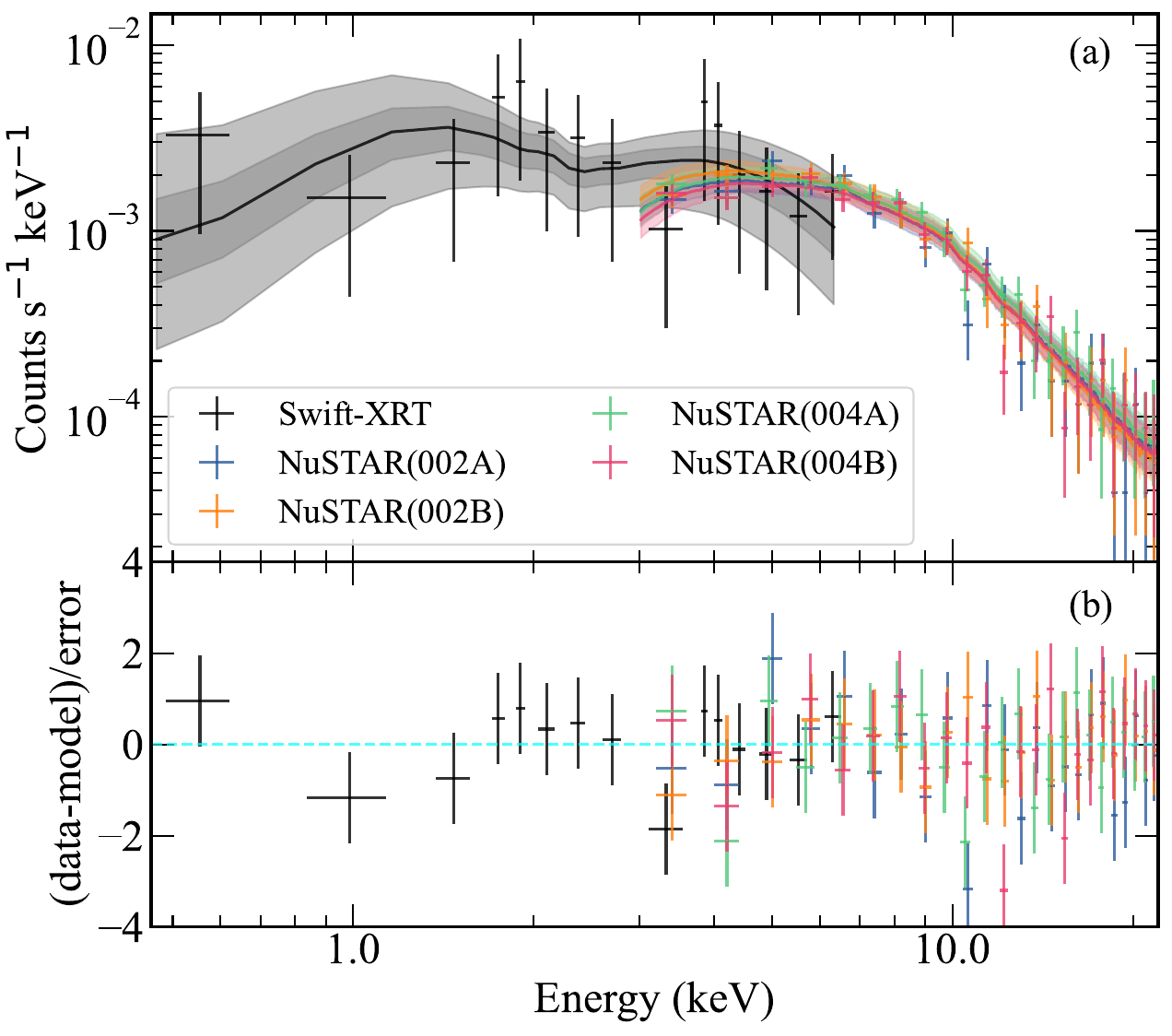}\\
\includegraphics[width=1.0\columnwidth]{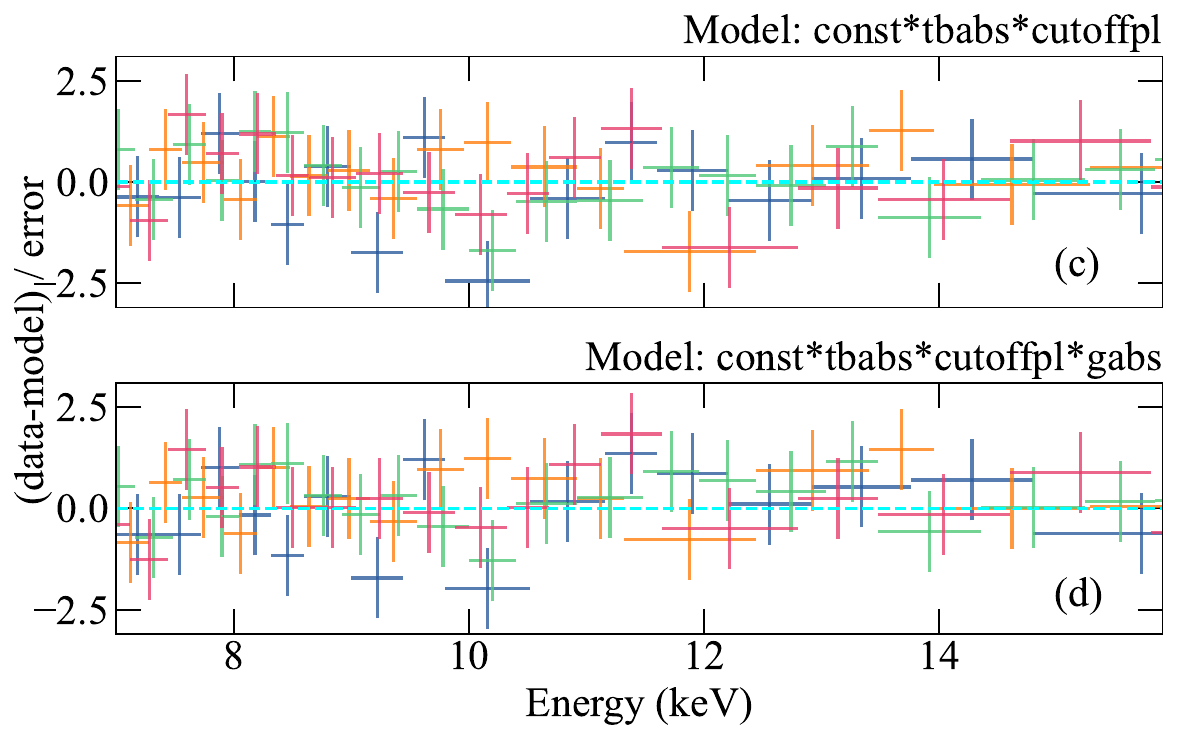}
\caption{Spectral fit of \src using the \swift/XRT and the \nustar data. Panel a: the spectra along with the best-fit models. Panel b: corresponding residuals. All spectra are rebinned for visual clarity. Panels c,d: detailed view of the 7--16 keV residuals before and after adding a \texttt{gabs} component, respectively.
}\label{fig:spec_sw_nu}
\end{figure}

Considering that numerous BeXRB systems exhibit CRSFs, we explored adding an additional Gaussian absorption component \texttt{gabs} to the model. While this marginally improved the fit quality, the physical parameters remained poorly constrained.
The result was consistent across other continuum models tested, indicating that the potential feature does not strongly depend on the specific continuum choice. As shown in Figure~\ref{fig:spec_sw_nu}, the inclusion of the \texttt{gabs} component could lead to a marginal improvement in the residuals.
We evaluated the significance of the improvement through the Monte Carlo approach using \texttt{simftest}. Based on 10$^4$ iterations, the probability that the data are consistent with a model lacking a \texttt{gabs} component is 3.45\%, which corresponds to a significance of approximately 1.8$\sigma$.
To physically characterize the potential absorption line, we first allowed its width to be a free parameter in the \texttt{gabs} model. This resulted in a poorly constrained value of $\sigma = 2.1^{+6.1}_{-1.1}$ keV.
To address potential systematics, we regenerated spectra using different bins and found that the absorption feature remained present and its centroid energy stable within uncertainties. We also assessed the impact of background subtraction by varying the background region selections and scaling. The significance and best-fit parameters of the feature were not substantially altered, confirming that it is not an artifact of the background modeling or data grouping.

We subsequently combined \xspec with \bxa to investigate potential CRSFs. The analysis revealed that the inclusion of an additional \texttt{gabs} component provides a statistically significant model improvement according to the likelihood comparison ($\Delta \log(L) > 1$). However, the absorption width parameter $\sigma$ remains poorly constrained. Systematic fitting attempts with various fixed $\sigma$ values demonstrated that $\sigma = 2.0$ keV led to the maximum likelihood and Bayesian evidence, though the evidence did not indicate significant improvement over alternative choices.

\begin{figure}
\centering
\includegraphics[width=1.0\columnwidth]{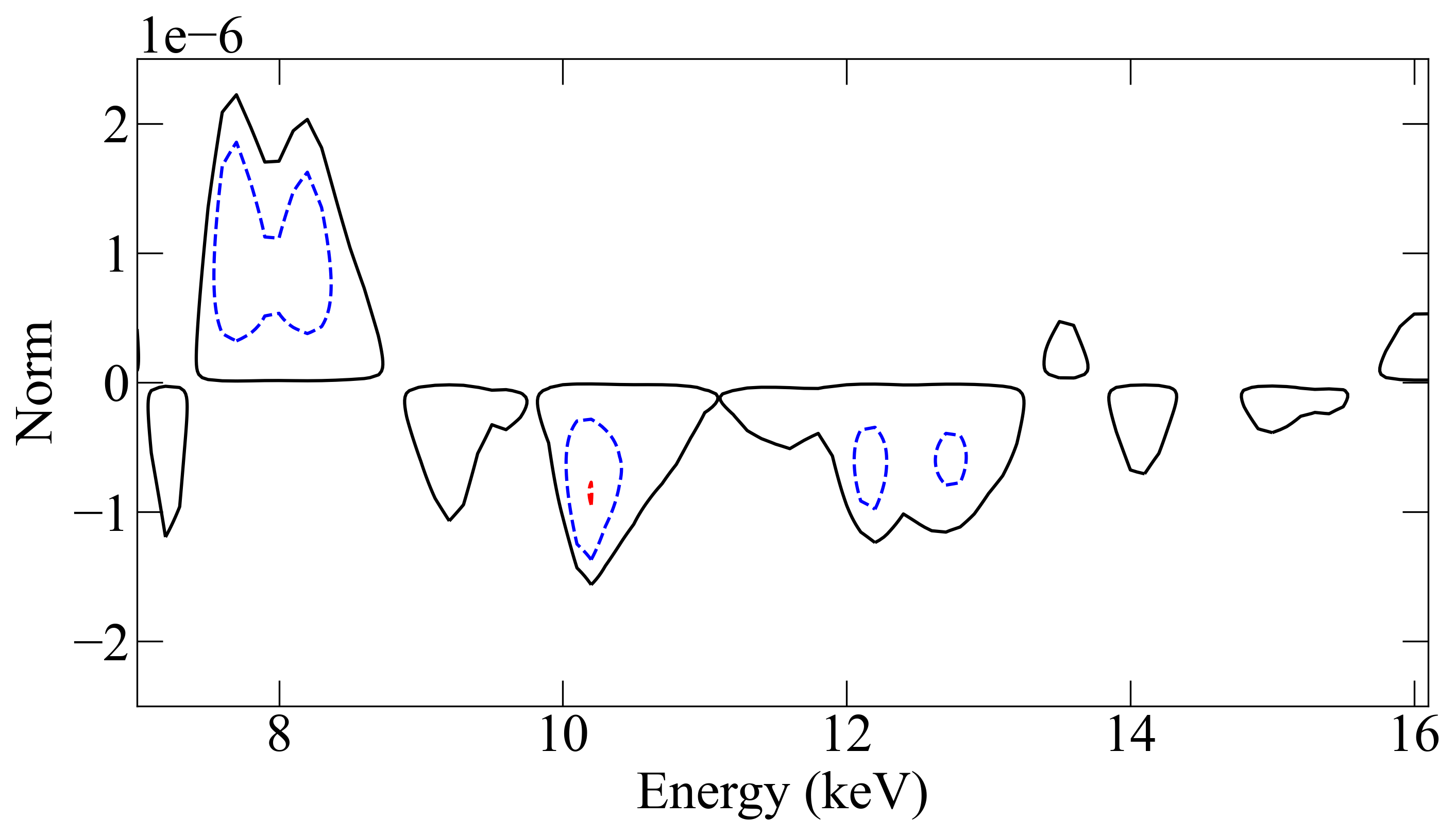}
\caption{Improvement of cstat $\Delta C$ contour plot with respect to the absorbed cutoff power-law model between 7--16 keV in the \nustar spectra from a blind search ($\Delta C=2.3$ in blue and $\Delta C=4.4$ in red). The contours in black indicate the baseline model reference level ($\Delta C=0.1$). }\label{fig:blsearch}
\end{figure}

Additionally, we performed a blind search for absorption features in the 7--16 keV following the method described in \citet{2010A&A...521A..57T} and \citet{2021NatAs...5..928S}. We defined the absorbed cutoff power law as the baseline model and added a narrow Gaussian line with a fixed width of the energy resolution of \nustar (40 eV) to search for the presence of possible emission or absorption features by stepping the centroid energy $E_0$ and normalisation N across the energy and normalisation grids. We recorded the improvement in the cstat $\Delta C$ relative to the baseline model at each grid point. The results of $\Delta C$ plotted in Figure~\ref{fig:blsearch} revealed marginally significant spectral features: a possible $\sim$8 keV emission structure and two absorption features at $\sim$10 keV and $\sim$12.3 keV, respectively. 
We noted that due to the statistical limitations of the data, it is not feasible to robustly assess the confidence levels of these features via $\chi^2$ based inference.
In \xspec, the $\sim$8 keV emission and $\sim$10 keV absorption structures could only be modelled with narrow Gaussian profiles when other parameters were frozen. We consider the $\sim$12.3 keV absorption structure to be more physically plausible and associated with the CRSF. Detailed discussion is provided in Section~\ref{sec:crsf_discussion}. 

Due to the large uncertainty and the low significance of the feature itself, as suggested by the \bxa analysis, we fitted the spectra with the absorption width fixed at $\sigma = 2$ keV and got a best fit with cstat = 701.74 for 937 degrees of freedom, though this improvement did not reach the 3-$\sigma$ confidence threshold. The fitting results are listed in Table \ref{tab:nu_para}, with all statistical 1-$\sigma$ errors for parameters.
The phase-resolved spectral measurements, in which photons were divided into several phase bins to extract spectra for fitting, showed that despite flux variations, spectral parameters such as the photon index did not evolve with phase.

\begin{table}
\centering
\caption{Model Parameters Comparison}
\label{tab:nu_para}
\renewcommand{\arraystretch}{1.5} 
\begin{threeparttable}
\begin{tabular}{llrr}
\hline\hline
Component & Parameter & {Value (\bxa)} & {Value (\xspec)} \\
\hline
\multicolumn{4}{l}{\texttt{const*tbabs*cutoffpl}} \\
\texttt{tbabs}    & $N_\mathrm{H}$        & 0.054     & 0.054       \\
\texttt{cutoffpl} & Phoindex  & $0.51_{-0.31}^{+0.26}$ & $0.61_{-0.31}^{+0.29}$  \\
         & $E_\mathrm{cutoff}$ (keV)  & $6.31_{-1.18}^{+1.45}$ & $6.03_{-1.16}^{+1.73}$  \\
         & norm ($10^{-5}$)     & $7.24_{-2.12}^{+2.53}$  & $8.95_{-2.41}^{+3.14}$ \\
\texttt{constant} & FPMB-002  & $1.11_{-0.08}^{+0.09}$ & $1.08_{-0.09}^{+0.09}$  \\
         & FPMA-004  & $1.03_{-0.07}^{+0.08}$ & $1.02_{-0.08}^{+0.08}$  \\
         & FPMB-004  & $1.05_{-0.08}^{+0.08}$ & $1.03_{-0.08}^{+0.09}$  \\
Total    & Statistic$^{a}$ & -6030.8            & 711.60/939              \\
\multicolumn{4}{l}{\texttt{const*tbabs*cutoffpl*gabs}} \\ 
\texttt{tbabs}    & $N_\mathrm{H}$        & 0.054    & 0.054       \\
\texttt{cutoffpl} & Phoindex  & $0.59_{-0.23}^{+0.25}$ & $0.40_{-0.33}^{+0.31}$  \\
         & $E_\mathrm{cutoff}$ (keV)  & $6.61_{-1.11}^{+1.52}$ & $6.03_{-1.15}^{+1.72}$  \\
         & norm ($10^{-5}$)     & $7.94_{-1.78}^{+2.53}$ & $6.52_{-1.95}^{+2.62}$  \\
\texttt{gabs}     & $E_\mathrm{CRSF}$ (keV)     & $12.13_{-0.79}^{+0.76}$ & $12.37_{-0.71}^{+0.72}$ \\
         & Sigma (keV)     & 2.0        & 2.0          \\
         & Strength  & $0.78_{-0.78}^{+0.85}$ & $2.31_{-0.75}^{+0.79}$  \\
\texttt{constant} & FPMB-002  & $1.12_{-0.09}^{+0.09}$ & $1.07_{-0.09}^{+0.09}$  \\
         & FPMA-004  & $1.03_{-0.07}^{+0.07}$ & $1.02_{-0.08}^{+0.08}$  \\
         & FPMB-004  & $1.05_{-0.07}^{+0.08}$ & $1.03_{-0.08}^{+0.08}$  \\
Total    & Statistic$^{a}$ & -6029.6            & 701.74/937              \\
\hline
\end{tabular}
\begin{tablenotes}
    \footnotesize
    \item (a) Value of the Bayesian evidence for \bxa and cstat/dof for \xspec.
\end{tablenotes}
\end{threeparttable}
\end{table}

The \nustar observations were performed merely one week after the \swift observation. However, the absorption-corrected flux estimated from the \nustar data was $9.05_{-0.98}^{+1.06} \times 10^{-13}$ \uergcms, only $\sim$1/3 of that measured by \swift and consistent with the flux estimation of the \ero data, which suggests that \src may have experienced a rapid flare around the time of the \swift observation.

\subsection{Long-term X-ray behaviour}

\begin{figure}
\centering
\includegraphics[width=1.0\columnwidth]{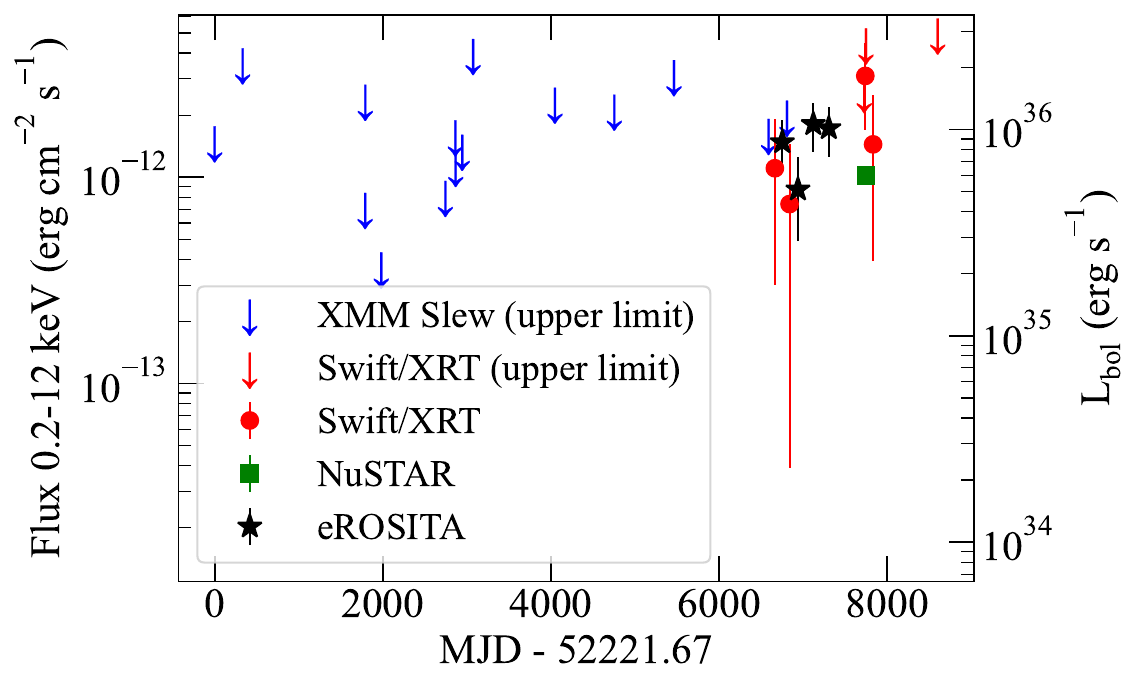}
\caption{Long-term X-ray light curve in the energy band 0.2--12.0 keV of \src. Corresponding bolometric X-ray luminosity values are calculated using the correction based on the \nustar data. Down-arrows mark 2-$\sigma$ upper limits.} \label{fig:uplmt_server}
\end{figure}

We investigated the long-term X-ray light curve of \src using the HIgh-energy LIght curve GeneraTor \citep[HILIGT;][]{2022A&C....3800529K, 2022A&C....3800531S}\footnote{\url{http://xmmuls.esac.esa.int/upperlimitserver/}}. For consistency and comparability we presented the 0.2--12 keV band results across all observations. We estimated the fluxes by assuming an absorbed power-law model with photon index of 0.5 and an absorption column density of 10$^{21}$ cm$^{-2}$. The \textit{ROSAT} data were excluded for the limited energy band. As shown in Figure~\ref{fig:uplmt_server}, \xmm slew observations contribute only upper limit fluxes. 
There were two \swift flux measurements in 2020 and one in April 2023, before and after our observation, respectively. However, these were derived from observations with short exposure times (< 100 s), resulting in relatively large uncertainties.

We calculated the 0.2--12 keV flux from the joint fit to the spectra of the two \nustar observations as described in Section~\ref{sec:spec_nustar} and then computed the broadband 0.1--100 keV bolometric X-ray luminosity $L_{\mathrm{X}}$. We calculated the bolometric correction to be $\sim$1.36 by taking the ratio of the bolometric luminosity to the luminosity in the 0.2--12 keV band, which was applied to estimate the bolometric X-ray luminosity in Figure~\ref{fig:uplmt_server}. The points plotted for \ero use the best-fit model described in Section~\ref{sec:ero_spec}, which is extrapolated to 0.2--12 keV and scaled by the average count rates during each eRASS compared to the total average count rate.

\subsection{OGLE monitoring of the optical counterpart}

\begin{figure}
\centering
\includegraphics[width=1.0\columnwidth]{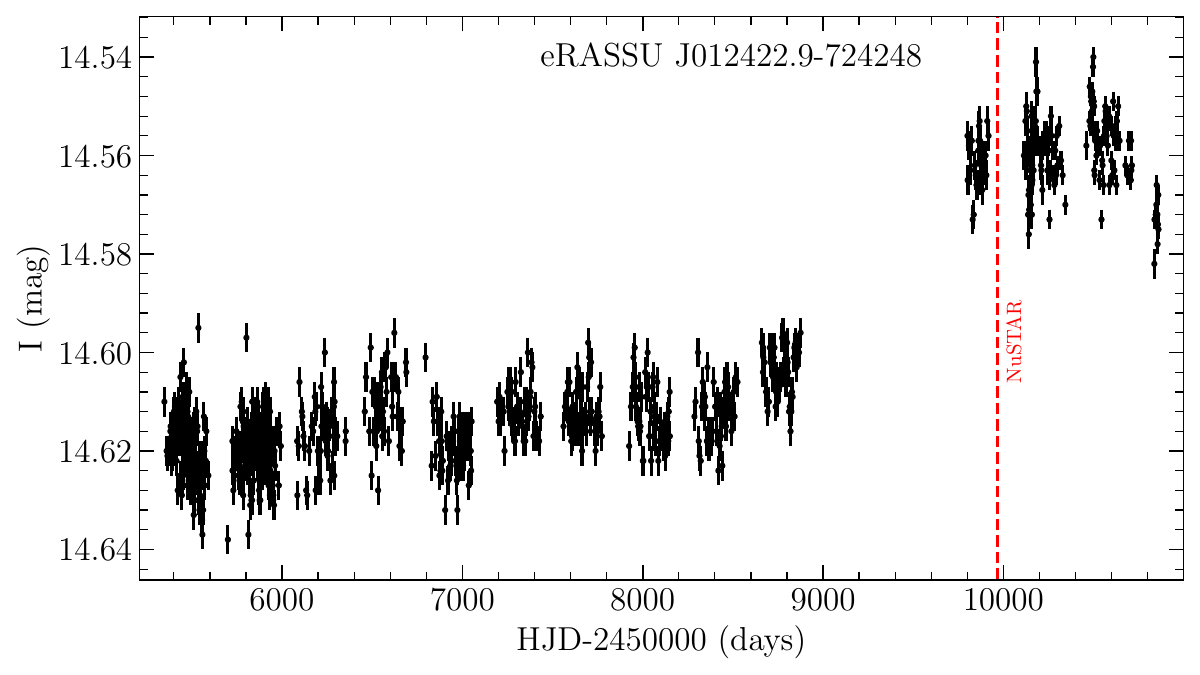}
\caption{OGLE-IV $I$-band light curve of the optical counterpart \osrc from 2010-05-31 to 2020-01-27 and 2022-08-09 to 2025-07-03. The long gap was caused by Covid-19. The red dashed line marks the epoch of the \nustar observation.}\label{fig:ogle_lc}
\end{figure}

We used the OGLE $I$-band light curve to investigate the optical long-term variability. As shown in Figure~\ref{fig:ogle_lc}, OGLE observed the optical counterpart of \src for about 15 years with a nearly 3-year gap caused by Covid-19. Overall, the amplitude of \osrc remains stable on timescales of several months, exhibiting variations of less than $\pm$0.02 mag in the $I$-band.
However, after the Covid gap the source brightness had increased by about 0.05 mag. 
We verified that this increase is real by checking a nearby star with similar brightness, which shows a flat light curve.
We searched for periodic signals from \osrc within the typical range of orbital periods found in BeXRBs \citep{2016A&A...586A..81H, 2022A&A...662A..22H, 2025A&A...694A..43T}. To suppress signal on long timescales due to the brightness increase, we adjusted the average I magnitude of the first part of the light curve to that of the second part.

\begin{figure}
\centering
\includegraphics[width=1.0\columnwidth]{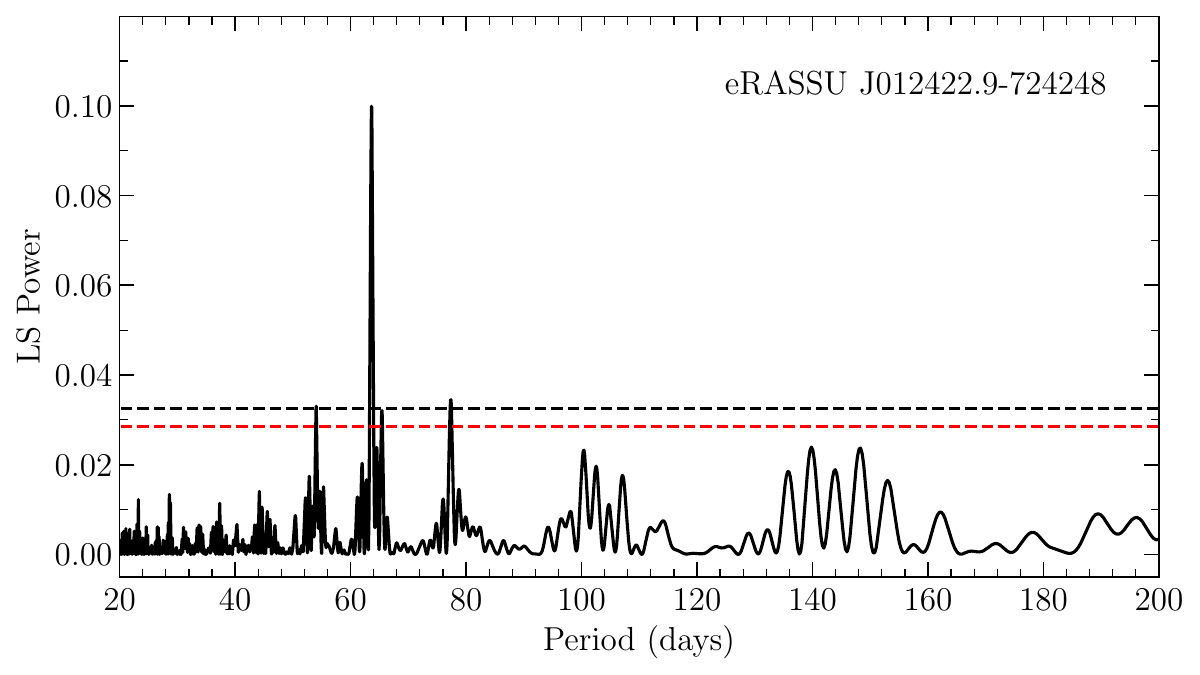}
\caption{The Lomb-Scargle periodogram of \src/\osrc obtained from the OGLE-IV $I$-band light curve. The dashed red and black lines mark the 95\% and 99\% confidence levels.}\label{fig:ogle_ls}
\end{figure}

\begin{figure}
\centering
\includegraphics[width=1.0\columnwidth]{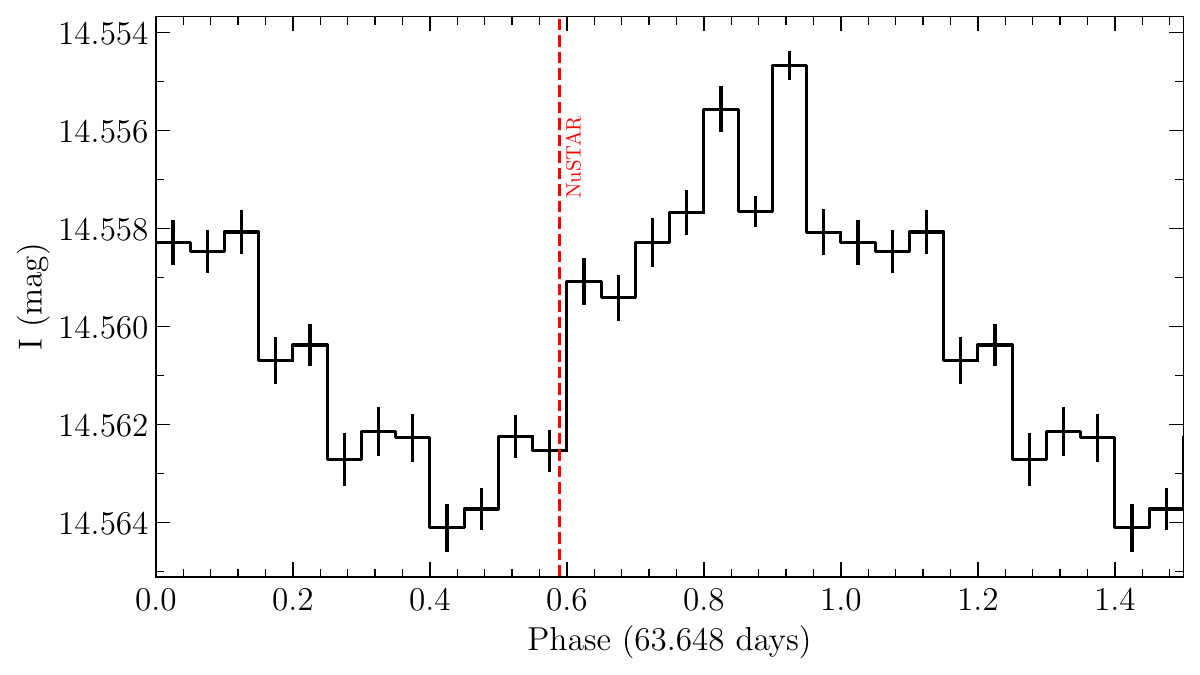}
\caption{OGLE $I$-band light curve folded with a period of 63.648\,days. The red dashed line marks the phase of the \nustar observation.}\label{fig:ogle_profile}
\end{figure}

We then produced the Lomb-Scargle periodogram from the adjusted light curve and searched for periodic variations. A significant signal at 63.648$\pm$0.002\,days (1$\sigma$ statistical error) was revealed, as shown in Figure~\ref{fig:ogle_ls}, which indicates the orbital period of the system.
Using the derived period, we folded the adjusted OGLE light curve, which reveals a broad single-peaked structure as shown in Figure~\ref{fig:ogle_profile}.

To investigate the stability of the period, we also analysed the OGLE-III light curve and divided the OGLE-IV light curve into two parts, before and after the Covid break. The strongest peaks in the Lomb-Scargle periodograms are recovered, but at slightly different periods (63.754$\pm$0.005\,days, 63.811$\pm$0.003\,days and 
63.924$\pm$0.010\,days for OGLE-III, the early and the later part of the OGLE-IV light curve, respectively).

\subsection{LCO/FLOYDS spectrum}
\begin{figure*}
\centering
\includegraphics[width=2.1\columnwidth]{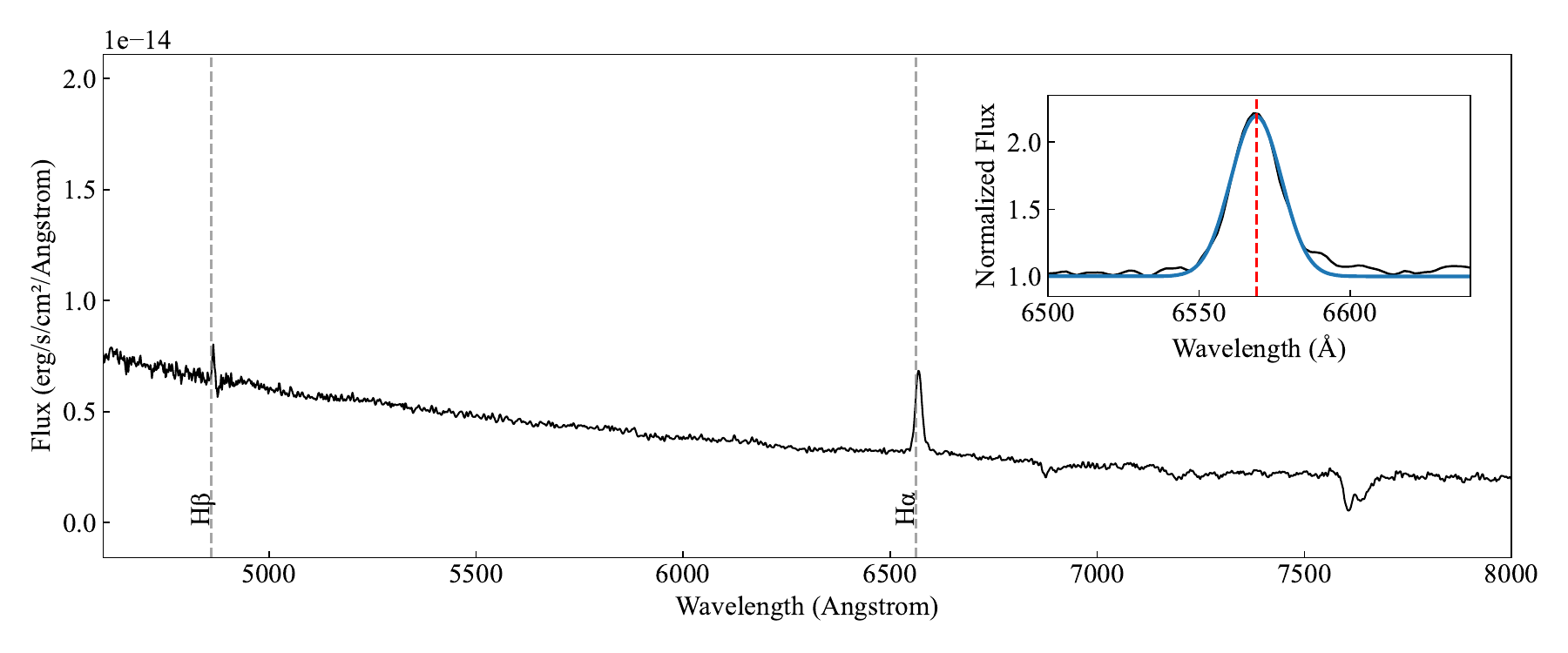}
\caption{Flux-calibrated optical spectrum of \src obtained using FLOYDS on LCO. 
The main panel shows the observed spectrum in physical units. The inset displays the continuum-normalized H$\alpha$ region, where the emission line is well described by a single-peaked Gaussian profile.\label{fig:lco_spec} }
\end{figure*}

To confirm the nature of the optical counterpart of \src as Be star, we investigated the optical spectrum obtained by the FLOYDS on LCO, which is presented in Figure~\ref{fig:lco_spec}. 
We analysed the Balmer series H$\alpha$ and H$\beta$ lines. The spectrum shows a strong H$\alpha$ emission line typical as seen from BeXRBs with a measured equivalent width of $-25.0\pm0.2$ \angstrom and a weaker H$\beta$ line with an equivalent width of $-1.2\pm0.9$ \angstrom.

\section{Discussion}\label{sec:discuss}

The MCs are connected by the Magellanic Bridge, which is believed to form tidally and draw stars and gas from LMC and SMC \citep{1990AJ.....99..191I, 2013A&A...551A..78B}. The Magellanic bridge is known to host several X-ray binary systems with a spatial distribution denser toward the Wing of the SMC \citep{2005A&A...435....9K, 2010MNRAS.403..709M, 2014MNRAS.444.3571S}, This indicates a different star formation history from that in the MCs.

Based on the broadband timing and spectral analysis using X-ray data from \ero, \swift, and \nustar, and optical data from OGLE and LCO, we report the discovery of the spin and orbital period of the system, and the identification of the optical counterpart as a Be star. This allows us to identify the neutron star BeXRB nature of \src.

\subsection{Periodicities}

The analysis of \nustar light curves reveals the X-ray pulsations of \src, which identifies a neutron star as the compact object in the system. The spin period of $341.71\pm 0.04$~s puts it in the slower rotating subpopulation of accreting pulsars, which is suggested to be associated with electron-capture supernovae or an advection-dominated accretion flow around the neutron star \citep{2011Natur.479..372K, 2014ApJ...786..128C}. Similar to SXP 265 \citep{2014MNRAS.444.3571S}, it is further supported by the relatively stable low-level X-ray flux observed in the \ero long-term light curve.
The orbital period of $\sim$63.65\,days obtained from OGLE optical observations places \src near the top region of the Corbet spin period - orbital period diagram (\citealt{1984A&A...141...91C}; \citealt{2009IAUS..256..361C}; and \citealt{2023A&A...669A..30M} for a recent version). In the Corbet diagram, BeXRBs with comparable orbital periods typically show shorter spin periods, and the source is more consistent with the population of such systems in the SMC.
The period derived from the OGLE light curve is not fully stable over the monitoring interval. Such variations are commonly observed from BeXRBs and interpreted as long-term changes in the circumstellar disc \citep{2011MNRAS.413.1600R}.

The pulse profile obtained from \nustar data shows a complicated energy-dependent shape, a characteristic commonly observed in numerous X-ray pulsars \citep[e.g.][]{1988ApJ...331..313D, 2002ApJ...581.1293R, 2008A&A...492..511K, 2025MNRAS.536.1357Y}. 
The 3--7 keV soft-band pulse profile shows some narrow dips, which could be caused by the partial obscuration of emitted radiation by matter in accretion streams \citep[e.g.][]{2008A&A...489..327H, 2013A&A...558A..74V, 2023A&A...669A..30M}. While heat maps from two \nustar observations suggest possible energy-dependent profile variations, the overall pulse shape shows no statistically significant differences between them.

\subsection{A tentative CRSF} \label{sec:crsf_discussion}

The broadband X-ray spectrum shows typical characteristics of a HMXB and is well described by an absorbed power law with a high-energy cutoff. 
The spectral fitting results suggest tentative evidence for an absorption feature near $\sim$12.3 keV, which, if confirmed as a CRSF, would allow an estimate of the magnetic field strength of the neutron star \citep{1978ApJ...219L.105T, 1981ApJ...251..288N,2019A&A...622A..61S}. 
The blind search provides two absorption structures. Although the most significant feature is located at $\sim$10 keV, attempts to fit it reveal only a very narrow structure. The best-fit width parameter $\sigma$ is much lower than the characteristic width expected for a CRSF which could be $\sim$1--2 keV using the temperature of the electrons $kT_e \sim$ 3.2 keV estimated with a \texttt{nthcomp} model for a self-emitting atmosphere \citep{1985ApJ...298..147M, 2009A&A...508..395R}. We therefore focused on the feature at $\sim$12.3 keV for the CRSF interpretation.

Given the limited statistical significance, the physical interpretation remains speculative. The two absorption features could, for instance, be manifestations of a single CRSF with an asymmetric profile, a phenomenon observed in sources like V\,0332+53 and linked to the relativistic cross-section or asymmetric line wings \citep{2005ApJ...634L..97P, 2025A&A...694A.316D}.  
Given that shape changes of pulsed fraction are known to be connected with some characteristics such as the CRSF and the Fe line \citep[e.g.][]{2009A&A...498..825F, 2009AstL...35..433L, 2023A&A...677A.103F}, the timing properties provide ancillary support as as the morphology of the pulsed fraction shows a broader drop at 9--12 keV.

Therefore, we identify a tentative CRSF at $\sim$12.3 keV, but its detection is not statistically robust.
Observations during future outbursts may help to improve the measurements.
Based on the CRSF energy of 12.13 keV inferred from the \bxa analysis, the magnetic field strength is estimated to $\sim 1.4 \times 10^{12}$ G following the 12-B-12 relation \citep{2007A&A...472..353S}.

\subsection{Long-term variability}

The long-term behaviour of \src shares some features with persistent BeXRB systems such as X Per and LS I +61 235 \citep{1998A&A...330..189H, 1998A&A...335..587H, 1999MNRAS.306..100R}. Although the X-ray luminosity of \src is higher than that of typical persistent low-luminosity systems like X Per, this similarity is further supported by the low cut-off energy and the absence of significant phase-dependent spectral variations in X-rays.
However, it is noted that the optical data of the source exhibits two states with different magnitudes despite the modest magnitude change of $\Delta I \sim 0.05$. Interestingly, during the optical high state there are also clues of enhanced X-ray activity, with detected flux changes of $\sim$30\% within one week according to the data of \swift and \nustar.

The $I$-band magnitude variations suggest changes in the circumstellar disk of the Be star occurring over a long timescale. Given that accretion onto the neutron star in BeXRBs primarily originates from the disk, X-ray and optical emissions typically show correlations on longer timescales \citep{2015A&A...574A..33R}. Besides, when the neutron star traverses the disk and triggers intense outbursts, a structural change may cause sustained brightening episodes \citep[e.g., RX J0520.5-6932;][]{2025MNRAS.536.1357Y}. However, despite our X-ray monitoring coverage during the gap of OGLE data, the possibility of some missed X-ray outbursts cannot be entirely excluded.

\section{Summary and Conclusions}

We report the detection of \src in the Magellanic Bridge, near the Eastern Wing of the SMC. The source was consistently detected in all four \ero all-sky surveys (eRASS1$-$4). Its X-ray position coincides with an early-type star identified as \osrc. 
Optical spectroscopy obtained with FLOYDS at LCO confirms strong H$\alpha$ emission from the companion star, with an equivalent width of $-25.0\pm0.2$ \angstrom, consistent with the characteristics of classical Be stars.

Timing analysis of the combined data from two \nustar observations reveals a spin period of $341.71\pm 0.04$ s, which confirms its identity as a member of the subpopulation of slower X-ray pulsars.
We found a periodic signal at 63.65\,days in the optical data of OGLE, which we suggest to be the binary orbital period. The modulation, which originates from the motion of the neutron star around its companion, is consistent with the behaviour of other BeXRBs in the MCs.

Broadband X-ray spectral analysis using \ero, \swift and \nustar data reveals that the spectrum is well characterized by an absorbed power law with a high-energy cutoff. In the \nustar spectrum a tentative CRSF is identified at $\sim$12.3 keV, which is marginally revealed by the blind search.

The long-term behaviour of the source appears to show persistent characteristics with the X-ray luminosity at a few $10^{35}$ \uergs, which suggests that \src can be a new  member of the class of persistent BeXRBs. Optical data reveal two distinct brightness states ($\Delta I \sim 0.05$) indicating long-term changes of the circumstellar disk.

\section*{Acknowledgements}

We acknowledge funding support from the National Natural Science Foundation of China (NSFC) under grant No. 12433005.
This work is based on data from \ero, the soft X-ray instrument aboard SRG, a joint Russian-German science mission supported by the Russian Space Agency (Roskosmos), in the interests of the Russian Academy of Sciences represented by its Space Research Institute (IKI), and the Deutsches Zentrum für Luft- und Raumfahrt (DLR). The SRG spacecraft was built by Lavochkin Association (NPOL) and its subcontractors, and is operated by NPOL with support from the Max Planck Institute for Extraterrestrial Physics (MPE). The development and construction of the \ero X-ray instrument was led by MPE, with contributions from the Dr. Karl Remeis Observatory Bamberg \& ECAP (FAU Erlangen-Nuernberg), the University of Hamburg Observatory, the Leibniz Institute for Astrophysics Potsdam (AIP), and the Institute for Astronomy and Astrophysics of the University of Tübingen, with the support of DLR and the Max Planck Society. The Argelander Institute for Astronomy of the University of Bonn and the Ludwig Maximilians Universität Munich also participated in the science preparation for \ero. The \ero data shown here were processed using the eSASS software system developed by the German \ero consortium.
This research has made use of data from the \nustar mission, a project led by the California Institute of Technology, managed by the Jet Propulsion Laboratory, and funded by the National Aeronautics and Space Administration. Data analysis was performed using the \nustar Data Analysis Software (NuSTARDAS), jointly developed by the ASI Science Data Center (SSDC, Italy) and the California Institute of Technology (USA).
The LCO observations have been made possible by the support of the Deutsche Forschungsgemeinschaft (DFG, German Research Foundation) under Germany’s Excellence Strategy-EXC-2094-390783311. 
HNY acknowledges support from China Scholarship Council (no. 202310740002). GV acknowledges support from the H.F.R.I. through the project ASTRAPE (Project ID 7802). AU acknowledges support from the “Copernicus 2024 Award” of the Polish FNP and German DFG agencies. The OGLE project has received funding from the Polish National Science Centre grant OPUS-28 2024/55/B/ST9/00447 to AU. LD acknowledges funding from the Deutsche Forschungsgemeinschaft (DFG, German Research Foundation) - Projektnummer 549824807.


\section*{Data Availability}

The data presented in the tables and figures of the paper are available upon reasonable request. The X-ray data are available through the High Energy Astrophysics Science Archive Research Center: \url{heasarc.gsfc.nasa.gov}.




\bibliographystyle{mnras}
\bibliography{012422_ref} 








\bsp	
\label{lastpage}
\end{document}